\documentclass[a4paper, amsfonts, amssymb, amsmath, reprint, showkeys, nofootinbib, twoside]{revtex4-1}
\usepackage[english]{babel}
\usepackage[utf8]{inputenc}
\usepackage[colorinlistoftodos, color=green!40, prependcaption]{todonotes}
\usepackage{amsthm}
\usepackage{xcolor}
\usepackage{graphicx}
\usepackage[T1]{fontenc}
\usepackage[pdftex, pdftitle={Article}, pdfauthor={Author}]{hyperref} 
\begin{document}

\title[TTP]{Thermodynamics of a transient phantom scenario}

\author{Miguel Cruz$^a$}
\email{miguelcruz02@uv.mx}

\author{Samuel Lepe$^b$}
\email{samuel.lepe@pucv.cl}

\affiliation{$^a$Facultad de F\'{\i}sica, Universidad Veracruzana 91097, Xalapa, Veracruz, M\'exico \\
$^b$Instituto de F\'{\i}sica, Facultad de Ciencias, Pontificia Universidad Cat\'olica de Valpara\'\i so, Avenida Brasil 2950, Valpara\'iso, Chile 
}

\begin{abstract}
This work is devoted to the thermodynamics description of a phantom scenario proposed previously by the authors. The presence of negative chemical potential is unavoidable if we allege for a well defined thermodynamics framework since the cosmological model passages from phantom stage at present time to a future de Sitter evolution. As noted earlier in other works, we find that the negativity of the chemical potential is necessary to save phantom dark energy from thermodynamics inconsistencies.  
\end{abstract}


\date{\today}

\maketitle

\section{Introduction}
\label{sec:intro}

The current crisis of the standard cosmological model (dubbed as $\Lambda$CDM) is hard to brush off. One of the most worrying problems, among other, of $\Lambda$CDM is the strong discrepancy in the predictions for the value of the Hubble constant ($H_{0}$) arising from the use of early-time and local measurements. These inconsistencies usually termed as {\it tensions} are not attributable anymore to systematic errors due to the development of the data collection and analysis, which is carried out by different collaboration groups. The precise determination of $H_{0}$ is important since establishes the rate of expansion of the universe and is useful to estimate its age, as is well known. A complete and very illustrative discussion about the discrepancies of the $H_{0}$ value can be found in Ref. \cite{riess}. For some sectors of the cosmological community the $H_{0}$ value problem in $\Lambda$CDM is evidence of new physics and reveals the necessity of more theoretical progress in order to describe the evolution of the universe more accurately, i.e., cosmological models beyond $\Lambda$CDM seem to be required; see for instance Refs. \cite{valentino, new} where a list of promising cosmological possibilities to solve the $H_{0}$ value problem is provided in each case. It is worthy to mention that in both aforementioned reviews, phantom dark energy is considered as a viable option to face some problems of the standard cosmological model. 

In Ref. \cite{vagnozzi1} it was explicitly shown that the consideration of non-standard values for the dark energy parameter state $\omega$ (phantom case) and for the effective number of relativistic species, usually termed as $N_{\mathrm{eff}}$; is enough to address the $H_{0}$ tension, this provides a new perspective for the model-building in cosmology. 

Additionally, in Ref. \cite{vagnozzi2} low redshift data as BAO and SNeIa slighly favored the phantom scenario over other effects in a string theory inspired cosmological model. We also refer the reader to Ref. \cite{beyond}, where deviations from $\Lambda$CDM are explored.  

Motivated by the prevailing tendency in the search for new cosmological models beyond the standard model, in a previous work we proposed the following form for the dark energy density in an homogeneous and isotropic universe, in the usual units, $8\pi G=c=k_{B}=1$, \cite{transient22}
\begin{equation}
\rho_{de}=3\alpha^{2}H^{4}, \label{eq:friednovel}
\end{equation}
with $\alpha > 0$ and $H$ being the Hubble parameter.  Then from the Friedmann constraint, $3H^{2} = \rho_{m}+3\alpha^{2}H^{4}$, the normalized expansion rate can be written as follows in terms of the redshift $z$
\begin{equation}
    E(z) = \left[\frac{1+\sqrt{1-\lambda(1+z)^{3}}}{2(1-\Omega_{m,0})} \right]^{1/2},\label{eq:rate}
\end{equation}
where the value of the parameter $\alpha$ is fixed by demanding that $E(z=0)=1$, i.e., the normalization condition is imposed at present time. From the previous equation one can write
\begin{equation}
\rho_{de}\left( z\right) = 3\left[ 1-\Omega_{m,0} \right] \left(\frac{1+\sqrt{1-\lambda \left( 1+z\right)^{3}}}{2\left[ 1-\Omega_{m,0} \right]}\right) ^{2}H^{2}_{0}, \label{eq:c1}
\end{equation}
being $\Omega_{m,0}$ the fractional energy density associated to the matter sector at present time, $H_{0}$ is the Hubble constant and we have defined the constant parameter, $\lambda := 4\left[ 1-\Omega_{m,0} \right]\Omega_{m,0}$, we refer the reader to Ref. \cite{transient22} for more details in the cosmological model.\\

The expression for the energy density given in equation (\ref{eq:friednovel}) was inspired from some different scenarios in cosmology. For instance, in Ref. \cite{salgado1} an energy density of the form given in (\ref{eq:friednovel}) appears naturally in the cosmology of the Einstein-Gauss-Bonnet gravity in 5D. On the other hand, the rate of expansion in the inflationary stage is modified by a term proportional to $H^{4}$ when entropic forces are included into the description, see Ref. \cite{entropic}, in both aforementioned cases the emergence of such a term is simply considered as a correction contribution to the cosmic evolution. However, we must bring to mind that the proportionality to $H$\footnote{A relation of the type $\rho_{de} \propto H$, has been also deeply explored in the holographic description of the universe when the Hubble scale is considered as characteristic length, see for instance the references (and references therein) given in \cite{holo}.} for the dark energy density is not unfamiliar and has been widely studied, such kind of dependence appeared for the first time in QCD \cite{ghost2}, where the magnitude of the vacuum energy is corrected by the contribution of some ghost fields given as $\rho \propto H \Lambda^{3}_{QCD}$, being $\Lambda_{QCD}$ the QCD mass scale. This is known as {\it ghost dark energy}. This latter result provides the right magnitude for the dark energy that explains the current accelerated expansion, see references given in \cite{qcd} for a complete guide on the topic, some of those works also considered minimal extensions for the ghost dark energy model by including a first derivative of $H$. 

In the context of higher curvature corrections to General Relativity, a $H^{4}$-term appears in the {\it Cosmological Einsteinian Cubic Gravity} and again such kind of term is interpreted as a correction since modules the evolution equation of the cosmic acceleration, the action of such gravity model is the following \cite{higher}
\begin{equation}
    S=\int d^{4}x\sqrt{-g}\left[\frac{1}{2\kappa}(R-2\Lambda)+\beta (\mathcal{P}-8\mathcal{C}) \right],
\label{eq:actioncubic}
\end{equation}
being $\mathcal{P}$ and $\mathcal{C}$ linear combinations of cubic Lagrangians of the form $R_{abcd}R^{abc}{}_{e}R^{de}$, $R_{abcd}R^{ac}R^{bd}$, $R_{ab}R^{ab}R$; to mention some. As shown in Ref. \cite{saridakiscubic}, a generalization of the form $\mathcal{P}\rightarrow f(\mathcal{P})$ in action (\ref{eq:actioncubic}) with $f$ being an arbitrary function of the cubic Lagrangians, can induce the emergence of terms up to sixth order in powers of $H$, in this case such terms can also be interpreted as corrections on the evolution equation of the cosmic acceleration. See also Ref. \cite{cisternaquartic}, where for the cosmological scenario is obtained again the arising of terms with higher powers of $H$ when the Einsteinian cubic gravity is extended to the quartic case. In the scenarios given above the late time cosmic acceleration is obtained in purely geometric terms. For our case the cosmic expansion is also characterized by only geometric terms since will be driven entirely by the energy density (\ref{eq:friednovel}).\\ 

Below we highlight the main features found in \cite{transient22} for a dark energy sector described by the energy density (\ref{eq:friednovel}) :
\begin{itemize}
\item the parameter state associated to the energy density (\ref{eq:friednovel}) can be written in terms of the coincidence parameter, $r:= \rho_{m}/\rho_{de}$, as follows
\begin{equation}
\omega_{de}(z) = -1 - 2\left(1+\omega_{m} \right)\frac{r(z)}{1-r(z)}, \label{eq:c2}
\end{equation}
the value $\omega_{m}=0$ will characterize the dark matter sector, as usual. Therefore, this dark energy model describes an early phantom scenario at the beginning of the dark energy dominance epoch\footnote{This epoch began at the recent past and we denote such beginning with the redshift value $z_{s}$.} which takes place at $r=1$, at this stage $\omega_{de}$ diverges. On the other hand, the explicit expression for the coincidence parameter is given as
\begin{equation}
r\left(z\right) =\frac{\lambda \left( 1+z\right) ^{3}}{\left( 1+\sqrt{1-\lambda \left(1+z\right)^{3}}\right)^{2}}. \label{eq:coincidence}
\end{equation}
Notice that at the far future, reached by the condition $z\rightarrow -1$, a de Sitter evolution characterized by $\omega_{de}(z\rightarrow -1)\rightarrow -1$ is recovered since $r(z\rightarrow -1)\rightarrow 0$; therefore the phantom behavior obtained in this dark energy model is transitory and the cosmic evolution in this model ends with a de Sitter phase.
\item The squared adiabatic sound speed is positive for this geometric dark energy model, which is signal of stability. 
\end{itemize}

In order to distinguish the cosmological implications of our model from those arising in a Big Rip scenario, we compute the energy 
\begin{equation}
    \mathcal{E}_{de}(z) = \rho_{de}(z)V(z)
\end{equation}
where $V(z) = V(z_{s})[(1 + z_{s})/(1 + z)]^{3}$ is the volume whose initial value is given by $V(z_{s})$ and $\rho_{de}(z)$ is written in Eq. (\ref{eq:c1}). It is worthy to mention that at the crossing the energy, $\mathcal{E}_{de}(z=z_{s})$, remains bounded and at present time ($z=0$) we have the condition $\mathcal{E}_{de}(z=0) > \mathcal{E}_{de}(z=z_{s})$. At the far future we have the limit, $\mathcal{E}_{de}(z\rightarrow -1)\rightarrow \infty$, and the divergent behavior of the energy is due to the volume definition. The energy associated to (\ref{eq:c1}) is always an increasing function. 
On the other hand, in a Big Rip scenario the Hubble parameter diverges at some time in the future which is characterized by $t_{s}$, in general we write the Hubble parameter as, $H(t) = \beta/(t_{s}-t)$ with $\beta >0$, $3H^{2}(t) = \rho$ and $V(t) \propto a^{3}(t)$. In this case the energy is given as follows
\begin{equation}
    \mathcal{E}_{BR}(t) = 3\beta^{2}a^{3}_{0}(t_{s}-t_{0})^{3\beta}\left(\frac{1}{t_{s}-t}\right)^{2+3\beta}, \label{eq:ocho}
\end{equation}
where $t_{0}$ denotes the present time and the subscript zero means evaluation of quantities at $t_{0}$. As can be observed, the energy (\ref{eq:ocho}) is also an increasing function but diverges at the crossing, $t=t_{s}$, i.e., $\mathcal{E}_{BR}(t=t_{0}) < \mathcal{E}_{BR}(t=t_{s})$. This is a crucial difference, in our model the singularity in the energy is kicked away to the far future ($z=-1$) or infinite time and the singular behavior at the crossing $z=z_{s}$ is obtained only in the parameter state (pressure) $\omega_{de}(z)$ of the fluid.\\

In the following section, we focus on the thermodynamics description for the phantom dark energy model obtained from the energy density (\ref{eq:friednovel}) and we explore the role of the chemical potential in this thermodynamics construction.
 
\section{Temperature of dark energy and the role of chemical potential}
\label{sec:ii}

The appropriate thermodynamics description for the dark energy sector is still subject of investigation since our knowledge about the nature of this component of the universe is incomplete. Furthermore, if the fluid describing dark energy enters the phantom region, which is not ruled out by observations, some physical principles could be violated. For instance, the second law of thermodynamics is disobeyed if we appeal to the positivity of the absolute temperature of the fluid. This issue is solved with the introduction of a new thermodynamic degree of freedom with the appropriate sign, see references given in \cite{chemical} for a review; this additional degree of freedom is the chemical potential (usually denoted by $\mu$) and appears naturally in the Euler relation, which is written below for a single fluid, \cite{chemical, callen, maartens, reif}
\begin{equation}
sT=\left( 1+\omega \right) \rho -\mu n, \label{eq:euler}
\end{equation}
where $n$ is the particle number density, i.e., the number of particles in the fluid is $N:=nV$, being $V:=a^{3}=(1+z)^{-3}$, the volume. Besides $s$ is the entropy density thus we have $S:=sV$ and $T$ stands for the absolute temperature of the fluid. We will assume the most simple relation between energy density and pressure, we restrict ourselves to a barotropic EoS given as $p_{i} = \omega_{i}\rho_{i}$, where the subscript $i$ denotes the different components. Since the number of particles in the fluid remains constant then $n$ is conserved as well as the energy of the fluid, their balance equations are 
\begin{align}
    & \dot{n}+3Hn=0, \label{eq:nointn}\\
    & \dot{\rho}_{de}+3H\rho_{de}(1+\omega_{de})=0,\label{eq:nonint}
\end{align}
respectively, the dot stands for derivatives w.r.t. cosmic time. According to Eq. (\ref{eq:nonint}) we are considering the non-interacting fluids approach. Therefore the temperatures of the species evolve independently, for the matter sector we have $\dot{\rho}_{m}+3H\rho_{m}=0$ then $T_{m} =T_{0}$, with $T_{0}$ being a constant \cite{maartens}. As found in Ref. \cite{lepegrandon}, the temperature of one sector can be influenced by the other within the interacting scenario, which provides the possibility $T_{de,m} \propto \pm Q$, where $Q$ is the interacting term and its sign determines the direction of the energy flux interchanged between both species. In the interacting case we write \cite{interact} 
\begin{align}
    & \dot{\rho}_{m}+3H\rho_{m}=-Q,\\
    & \dot{\rho}_{de}+3H\rho_{de}(1+\omega_{de})=Q,
\end{align}
leading to a non-conservation of energy per specie but the total energy is conserved. For $Q=0$ we recover the non-interacting fluids scheme and the temperatures evolve normally.\\ 
Notice that if the positivity condition for the absolute temperature is relaxed in Eq. (\ref{eq:euler}), then the second law can still be fulfilled with a phantom fluid if the adequate sign for the chemical potential is considered. This approach differs considerably from the studies performed in \cite{chemical}, such relaxation in the sign of the absolute temperature is necessary in our scheme as we will comment later.

Negative absolute temperatures are not strange in nature, the appearance of this phenomena was reported in 1951 by Purcell and Pound as consequence of quantum physics in a system of nuclear spins which exhibited such characteristic when the applied magnetic field to the system was rapidly reversed \cite{nat1}. Physically the reciprocal temperature, $\theta :=1/T$, is defined as
\begin{equation}
    \theta = \frac{\partial S}{\partial U}, \label{eq:partial}
\end{equation}
being $U$ the internal energy of the system, in the fluid description we have $U=\rho V$. Then, the energy is transferred from a system with lower $\theta$ to one with a higher $\theta$ implying an increasing behavior for the total entropy according to (\ref{eq:partial}), thus, if we specialize to the case $T<0$ we obtain that a system with negative temperature is hotter than one with positive temperature. In this case the occupation distribution increases exponentially since, $P_{i} \propto e^{-E_{i}/T}$, and states with high energies are more occupied than those with low energies, in other words, the occupation is inverted and is a transient phenomenon only viable for systems with energy spectrum limited from above, the existence of both aforementioned properties in a system with negative absolute temperature is of interest for our purposes. A complete discussion for this kind of physical systems can be found in Ref. \cite{nat2}.

As can be seen, the usual expression for the Gibbs law \cite{maartens, callen, reif}
\begin{equation}
    nTd\left(\frac{s}{n}\right) = d\left(\frac{\rho}{n}\right) + pd\left(\frac{1}{n}\right)-\mu d\left(\frac{1}{n}\right),\label{eq:gibbs}
\end{equation}
leads to the adiabatic condition, when the conservation equations for $\rho$ and $n$ are used, then the entropy is a definite positive constant. There is not interchange of any kind between the dark energy fluid and other components, this is consistent with the single fluid description. In order to describe the dark energy fluid we insert Eq. (\ref{eq:c2}) into the Euler relation (\ref{eq:euler}), yielding
\begin{equation}
sT_{de}=-\left[ \left( \frac{2r}{1-r}\right) \rho_{de} +\mu n\right].
\end{equation}
Therefore we can compute the following expression for the temperature of dark energy by means of our previous results
\begin{equation}
T_{de}\left( z\right) =-\frac{N}{S}\left[ \frac{2\left( 1+z_{s}\right) ^{3}}{n\left( z_{s}\right) }\left( \frac{r\left( z\right) \left( 1+z\right) ^{-3}}{1-r\left( z\right) }\right) \rho_{de} \left( z\right) +\mu \right],\label{eq:nattemp}
\end{equation}
where $\rho_{de}(z)$ is given in Eq. (\ref{eq:c1}). Notice that our description is valid only from $z_{s}$ to $z=-1$, thus the initial value of the physical quantities is given at $z=z_{s}$, for instance from Eq. (\ref{eq:nointn}) the particle number density can be written as, $n(z) = n(z_{s})[(1+z)/(1+z_{s})]^{3}$. Given that the number of particles does not change it means that its initial value, $N=n(z_{s})V(z_{s})$, remains constant.\\

The energy density (\ref{eq:c1}) for the phantom fluid remains bounded all the time and tends to the maximum value $3H^{2}_{0}/(1-\Omega_{m,0})$ at $z=-1$; this is consistent with the requirement of energy spectrum limited from above for a system with negative absolute temperature.\\

$\bullet$ {\bf Chemical potential and the consistency for the de Sitter limit}\\

For null chemical potential we observe that in the limit $z\rightarrow -1$ one gets for the temperature given in (\ref{eq:nattemp}) (see equations (\ref{eq:c1}) and (\ref{eq:coincidence}))
\begin{equation}
    T_{de}(z\rightarrow -1) = -\frac{N}{S}\left[ \frac{3H^{2}_{0}\lambda \left( 1+z_{s}\right) ^{3}}{2n\left( z_{s}\right)\left(1-\Omega_{m,0} \right)}\right],
\end{equation}
which is a negative constant. According to our description, the limit $z\rightarrow -1$ leads to a de Sitter evolution, in such case the temperature must fulfill the condition $T(z=-1) = 0$, which is the expected result for a dark energy described by a constant energy density, see for instance \cite{lepegrandon, saridakis}, therefore we have an inconsistent behavior for the temperature with null chemical potential. However, if we consider the case $\mu = \mbox{constant} \neq 0$, one gets the following temperature at the far future
\begin{equation}
T_{de}\left( z\rightarrow -1\right) \rightarrow -\frac{N}{S}\left[ \frac{6\Omega
_{m,0}\left( 1+z_{s}\right) ^{3}H_{0}^{2}}{n\left( z_{s}\right)}+\mu \right],
\end{equation}
where the definition of $\lambda$ has been used. In this case we can specify the form of the chemical potential as follows
\begin{equation}
    \mu = - \frac{6\Omega_{m,0}\left( 1+z_{s}\right) ^{3}H_{0}^{2}}{n\left( z_{s}\right)}, \label{eq:chemlast}
\end{equation}
in order to recover the condition, $T(z=-1) = 0$. This expression for the chemical potential indicates us that at the crossing given by the value $z_{s}>0$, we obtain a non-null and bounded chemical potential. This differs from the phantom scenario discussed in \cite{saridakis}, where the chemical potential is null at the crossing. Finally, the form of the temperature can be penned as
\begin{widetext}
\begin{equation}
    T_{de}\left( z\right) = -\frac{2\left( 1+z_{s}\right) ^{3}}{n\left(
z_{s}\right) }\left( \frac{N}{S}\right) \left[ \left( \frac{r\left( z\right)
\left( 1+z\right) ^{-3}}{1-r\left( z\right) }\right) \rho_{de}\left( z\right)
-3\Omega_{m,0}H_{0}^{2}\right].
\label{eq:nega}
\end{equation}
\end{widetext}
Notice that at the crossing, $z_{s}$, the temperature exhibits a singularity since $r(z_{s})=1$. The recovery of a null temperature for the de Sitter limit is backed by the inclusion of chemical potential in the thermodynamics description, the value of the chemical potential is restricted to be negative and is given in terms of the cosmological parameters $\Omega_{m,0}$ and $H_{0}$. The values $\mu = 0$ and $\mu > 0$ do not satisfy the de Sitter evolution condition, $T=0$. Notice that the third law of thermodynamics \cite{reif} is obeyed in this cosmological model given since the cosmic evolution is adiabatic and the limit, $T(z\rightarrow -1)\rightarrow0$, is guaranteed with the inclusion of negative chemical potential. This is an interesting behavior since in such limit for the temperature the occupation distribution is expected to be restored, as mentioned above, a requirement for a system with negative absolute temperature is the transience of the inverted occupation distribution. The synergy between $\mu$ and $T$ it is not unknown, a generalization of the Tolman-Ehrenfest law given in \cite{relate1}, exhibits an intimate relation between $\mu$ and $T$ through a definite position dependent expression in a general fluid, a relation of this kind for the pair of thermodynamic variables $(\mu, T)$ was previously explored by Klein in \cite{relate2}. As shown above, the negativity condition for the chemical potential in this dark energy model is necessary to guarantee thermodynamic consistency, see for instance \cite{chemical}, where the sign of the chemical potential has an important role to save the phantom hypothesis from thermodynamics disaster. In this context, for cosmological dissipative fluids described by the framework of thermodynamics of irreversible processes within the causal (linear and non-linear) Israel-Stewart formalism, the inclusion of chemical potential is necessary in order to describe the phantom zone without entering into contradictions with the definition of temperature and entropy given by standard thermodynamics. In such case the negativity of the chemical potential is crucial to maintain the positivity of the temperature and entropy at the same time, i.e., we have $TS>0$ for $\mu <0$, thus from the latter condition the agreement of the model with the second law of thermodynamics is guaranteed; see Ref. \cite{chemical2}, where the resolution to the negative entropy or negative temperature problem in accelerating phantom universe was proposed at first time using dissipative fluids.\\

According to our results we can write equation (\ref{eq:chemlast}) simply as
\begin{equation}
    \mu = -g, 
\end{equation}
where we defined $g:=G/N$. Since the expression for the chemical potential is obtained from the Euler relation (\ref{eq:euler}) in the limit $z\rightarrow -1$, we can observe that $g$ corresponds to the Gibbs free energy per particle with $G = (1+\omega)\rho-sT$. Therefore the chemical potential is simply $g$ since we are describing a single fluid \cite{reif}, $\mu \neq g$ for two or more fluids or different species. Notice that in this case any change in $G$ denoted as $\Delta G$ is null, this behavior characterizes a state of equilibrium, i.e., the constituents of this dark energy fluid remain unaltered along the cosmic evolution.\\

\begin{figure}[htbp!]
\centering
\includegraphics[scale=0.65]{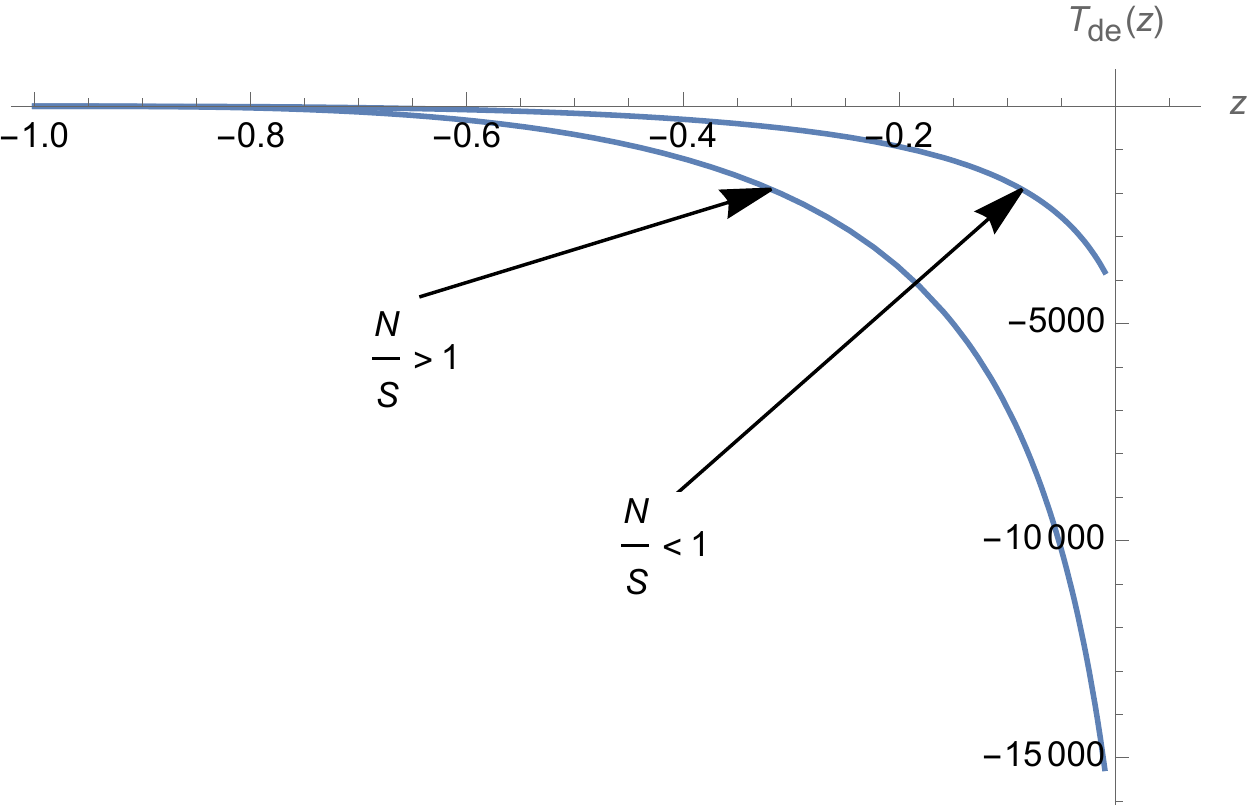}
\caption{Behavior of $T_{de}(z)$. For this plot we have considered $\Omega_{m,0}=0.3$ and $z_{s} \approx 0.06$, which results from the solution of the condition $r(z) = 1$ in Eq. (\ref{eq:coincidence}), the same value for $H_{0}$ was used in both cases.}
\label{fig:temperature}
\end{figure}

On the other hand, the specific heat (heat capacity) for the dark energy sector, can be computed from the relation $\partial \rho/\partial T = (\partial \rho/\partial z)(\partial T/\partial z)^{-1}$, using Eqs. (\ref{eq:c1}), (\ref{eq:coincidence}) and (\ref{eq:nega}). We obtain a regular behavior at $z=z_{s}$ for the specific heat and diverges at $z=-1$. This is an interesting behavior since the crossing to the phantom regime given at the past, $z=z_{s}$, does not represent a thermodynamic phase transition. As argued in Ref. \cite{saridakis}, a radical phase transition at the recent past of the cosmic evolution would have left observable cosmological imprints. However, the model exhibits a phase transition at the far future, characterized by the recovery of the de Sitter limit.

An interesting characteristic of negative temperature systems is their negative pressure, this can be seen from directly from the Euler relation (\ref{eq:euler}), $p = -\rho +sT +\mu n$, for $T<0$ and $\mu n >0$ we have $p<0$ if the term $\mu n$ is negligible with respect to the sum of the first and second terms, this characteristic was also discussed in \cite{nat2} for negative temperature states of motional degrees of freedom. In our case we have a negative chemical potential, then the phantom fluid has a very negative pressure which tends to a bounded value as $z\rightarrow -1$. See also Ref. \cite{nat3} and references therein for an interesting discussion on negative absolute temperatures.

From the cosmological point of view, in Ref. \cite{nat4} several interesting results were found for cosmological models that admit expanding solutions and negative absolute temperature, for instance, (i) these kind of solutions experience a phantom expansion and posses an attractor fixed point characterized by $T=0^{-}$ together with $\rho = \rho_{max}$ and $\omega=-1$, this corresponds to a de Sitter expansion; (ii) if the cosmic description begins with $T = -\infty$, the evolution to the vicinity of $T = 0^{-}$ is extremely rapid and (iii) the adiabatic sound speed could take large values (but positive) for a very short time interval. It is worthy to mention that the aforementioned characteristics are in agreement with those exhibited by our dark energy scenario with $\mu < 0$. Therefore, our construction appears to be consistent. A complete discussion around the negativity of the temperature for the phantom regime can be found in \cite{nat5, nat6}. 

To end this section, we comment about the cosmological implications of our model on the horizon thermodynamics. For a flat Friedmann-Lemaitre-Robertson-Walker universe the radius of the apparent horizon is given by \cite{apparent, cai}
\begin{equation}
    R_{A} = \frac{1}{H},
\end{equation}
and the temperature of the apparent horizon for an expanding universe is
\begin{eqnarray}
    \mathcal{T}_{A} &=& \frac{1}{2\pi R_{A}}\left( 1-\frac{\dot{R}_{A}}{2HR_{A}}\right) = \frac{T_{A}}{2}\left(1-q\right)\nonumber \\ 
    &=& \frac{T_{A}}{4}\left( 1-3\frac{\omega_{de}(z)}{1+r(z)}\right), \label{eq:Taomega}
\end{eqnarray}
where $T_{A} := 1/(2\pi R_{A})$ is the Cai-Kim temperature \cite{cai}, $q$ is the deceleration parameter defined as $1+q:=-\dot{H}/H^{2}$ and we have considered the acceleration equation $2\dot{H} + 3H^{2} = -p$ to write Eq. (\ref{eq:Taomega}), being $p$ the total pressure. As can be seen, the horizon temperature can be written in terms of the matter content of the universe; at the crossing we have, $\mathcal{T}_{A}(z=z_{s})=\infty$, and for phantom regime ($q<-1$ or $\omega_{de}<-1$) the temperature $\mathcal{T}_{A}$ is positive. The de Sitter case characterized by $q=-1$ or $\omega_{de}=-1$ leads to $\mathcal{T}_{A} = T_{A}$. Therefore, the positivity of temperature on the horizon is fulfilled for this cosmological model. For an infinitesimal time $dt$, the amount of energy crossing the apparent horizon due to the matter content of the universe is found to be \cite{apparent, cai, appa2}
\begin{equation}
    dE = -A(\rho + p)HR_{A}dt,\label{eq:flux}
\end{equation}
where $A=4\pi R^{2}_{A}$ is the area of the apparent horizon and the entropy is simply $S_{A} = A/4$. The heat flow $\delta Q$ is related to the amount of energy crossing the horizon in the following form by means of the Cai-Kim Clausius equation \cite{cai, appa2}, $\delta Q = -dE = \mathcal{T}_{A}dS_{A}$. Notice that we can calculate independently $\delta Q$ using the expression $\mathcal{T}_{A}dS_{A}$ and $dE$ from (\ref{eq:flux}). Using our previous results we can write
\begin{equation}
    \mathcal{T}_{A}dS_{A} = \frac{3\pi}{4}T_{A}\left(1-\frac{3\omega_{de}(z)}{1+r(z)}\right)\left(1+\frac{\omega_{de}(z)}{1+r(z)} \right)dt, \label{eq:heat1}
\end{equation}
where $dS_{A} = 2\pi \dot{R}_{A}dt$ and (\ref{eq:flux}) takes the form
\begin{equation}
    dE =-A(1+r(z))\left(1+\frac{\omega_{de}(z)}{1+r(z)} \right)\rho_{de}(z)HR_{A}dt. \label{eq:heat2}
\end{equation}
According to Eqs. (\ref{eq:heat1}) and (\ref{eq:heat2}), for the de Sitter case we obtain $\delta Q = \mathcal{T}_{A}dS_{A} =0$ and $dE=0$; for the phantom scenario one gets $\delta Q = \mathcal{T}_{A}dS_{A} < 0$ and $dE > 0$. This is consistent with the sign convention (positive heat out) $\delta Q = -dE$, which means that heat emitted by the system takes positive values and heat absorbed by the system takes negative values \cite{appa2}. The results shown above are in agreement with the phantom scenarios discussed in \cite{sariappa1, sariappa2} where the horizon thermodynamics is addressed.      

\section{Conclusions}
\label{sec:iv}
In this work we explored some thermodynamics aspects of a dark energy model previously proposed by the authors in Ref. \cite{transient22}, that exhibits a transient phantom stage. The cosmological model leads to a negative absolute temperature and fulfills some of the properties found in the literature for these kind of cosmological scenarios, being its most interesting behavior the evolution towards a de Sitter expansion at the far future. This characteristic is important since the de Sitter evolution represents a fixed point of any cosmological model with negative temperature and expanding solutions. 

In our description the presence of a negative chemical potential is essential in order to recover the null temperature for the dark energy fluid when the de Sitter evolution is reached. It was found that such chemical potential is written in terms of the cosmological parameters $(\Omega_{m,0},H_{0})$ and is simply the Gibbs free energy per particle since we are dealing with a single fluid, therefore no new arbitrary degrees of freedom are introduced in our scheme. Once again was shown that the chemical potential is useful to solve some pathologies at thermodynamics level of dark energy models that admit a phantom regime \cite{chemical}.

An open subject to explore in this kind of dark energy model is a test against observations in order to determine if the existent tensions within $\Lambda$CDM can be alleviated or worsened with the use of this theoretical framework. It is worthy to mention that in the recent past (in the vicinity of $z=0$, i.e., late times), the expansion rate described by (\ref{eq:rate}) is lowered with respect to $\Lambda$CDM; this is an interesting feature of our model since to address the $H_{0}$ tension, the most simple possibility at hand is the lowering of the expansion rate at late times by modifying the dark energy sector, see for instance \cite{vagnozzi2} and \cite{vagnozzi3} where a closer examination of the aforementioned condition is investigated for different cosmological models. We will discuss these crucial aspects of the model elsewhere.  

\section*{Acknowledgments}
MC work has been supported by S.N.I.I. (CONAHCyT-M\'exico). SL thanks P. Vargas (UTFSM) for his interesting comments on thermodynamics.


\begin{thebibliography}{0}%
\makeatletter
\providecommand \@ifxundefined [1]{%
 \@ifx{#1\undefined}
}%
\providecommand \@ifnum [1]{%
 \ifnum #1\expandafter \@firstoftwo
 \else \expandafter \@secondoftwo
 \fi
}%
\providecommand \@ifx [1]{%
 \ifx #1\expandafter \@firstoftwo
 \else \expandafter \@secondoftwo
 \fi
}%
\providecommand \natexlab [1]{#1}%
\providecommand \enquote  [1]{``#1''}%
\providecommand \bibnamefont  [1]{#1}%
\providecommand \bibfnamefont [1]{#1}%
\providecommand \citenamefont [1]{#1}%
\providecommand \href@noop [0]{\@secondoftwo}%
\providecommand \href [0]{\begingroup \@sanitize@url \@href}%
\providecommand \@href[1]{\@@startlink{#1}\@@href}%
\providecommand \@@href[1]{\endgroup#1\@@endlink}%
\providecommand \@sanitize@url [0]{\catcode `\\12\catcode `\$12\catcode
  `\&12\catcode `\#12\catcode `\^12\catcode `\_12\catcode `\%12\relax}%
\providecommand \@@startlink[1]{}%
\providecommand \@@endlink[0]{}%
\providecommand \url  [0]{\begingroup\@sanitize@url \@url }%
\providecommand \@url [1]{\endgroup\@href {#1}{\urlprefix }}%
\providecommand \urlprefix  [0]{URL }%
\providecommand \Eprint [0]{\href }%
\providecommand \doibase [0]{http://dx.doi.org/}%
\providecommand \selectlanguage [0]{\@gobble}%
\providecommand \bibinfo  [0]{\@secondoftwo}%
\providecommand \bibfield  [0]{\@secondoftwo}%
\providecommand \translation [1]{[#1]}%
\providecommand \BibitemOpen [0]{}%
\providecommand \bibitemStop [0]{}%
\providecommand \bibitemNoStop [0]{.\EOS\space}%
\providecommand \EOS [0]{\spacefactor3000\relax}%
\providecommand \BibitemShut  [1]{\csname bibitem#1\endcsname}%
\let\auto@bib@innerbib\@empty
\end{thebibliography}%


\begin{thebibliography}{99}
\bibitem{riess}
A.~G.~Riess, Nat.\ Rev.\ Phys. {\bf 2}, 10 (2020).

\bibitem{valentino}
E.~Di~Valentino, O.~Mena, S.~Pan, L.~Visinelli, W.~Yang, A.~Melchiorri, D.~F.~Mota, A.~G.~Riess and J.~Silk, Class.\ Quantum\ Grav. {\bf 38}, 153001 (2021).

\bibitem{new}
Jian-Ping~Hu and Fa-Yin~Wang, Universe {\bf 9}, 94 (2023).

\bibitem{vagnozzi1}
S.~Vagnozzi, Phys.\ Rev.\ D {\bf 102}, 023518 (2020).

\bibitem{vagnozzi2}
L.~Visinelli, S.~Vagnozzi and U.~Danielsson, Symmetry {\bf 11}, 1035 (2019).

\bibitem{beyond}
L.~Knox and M.~Millea, Phys.\ Rev.\ D {\bf 101}, 043533 (2020).

\bibitem{transient22}
M.~Cruz, S.~Lepe and G.~E.~Soto, Phys.\ Rev.\ D {\bf 106}, 103508 (2022).

\bibitem{salgado1}
F.~Gomez, S.~Lepe, V.~C.~Orozco and P.~Salgado, Eur.\ Phys.\ J.\ C {\bf 82}, 906 (2022).

\bibitem{entropic}
D.~A.~Easson, P.~H.~Frampton, G.~F.~Smoot, Int.\ J.\ Mod.\ Phys.\ A {\bf 27}, 1250066 (2012).

\bibitem{holo}
A.~Cohen, D.~Kaplan and A.~Nelson, Phys.\ Rev.\ Lett. {\bf 82}, 4971 (1999); S.~D.~H.~Hsu, Phys.\ Lett.\ B \textbf{594}, 13 (2004); V.~H.~C\'ardenas, M.~Cruz and S.~Lepe,	Phys.\ Dark\ Univ. {\bf 37}, 101122 (2022).

\bibitem{ghost2}
E.~Witten, Nucl.\ Phys.\ B {\bf 156}, 269 (1979); G.~Veneziano,
Nucl.\ Phys.\ B {\bf 159}, 213 (1979); C.~Rosenzweig, J.~Schechter and C.~G.~Trahern, Phys.\ Rev.\ D {\bf 21}, 3388 (1980); P.~Nath and R.~L.~Arnowitt, Phys.\ Rev.\ D {\bf 23}, 473 (1981); K.~Kawarabayashi and N.~Ohta, Nucl.\ Phys.\ B {\bf 175}, 477 (1980).

\bibitem{qcd}
F.~R.~Urban and A.~R.~Zhitnitsky, Phys.\ Lett.\ B {\bf 688}, 9 (2010); N.~Ohta, Phys.\ Lett.\ B {\bf 695}, 41 (2011); A.~R.~Zhitnitsky, Phys.\ Rev.\ D {\bf 92}, 043512 (2015); A.~O.~Barvinsky and A.~R.~Zhitnitsky, Phys.\ Rev.\ D {\bf 98}, 045008 (2018); B.~Holdom, Phys.\ Lett.\ B {\bf 697}, 351 (2011); A.~Yamamoto, Phys.\ Rev.\ D {\bf 90}, 054510 (2014); Rong-Gen~Cai, Zhong-Liang~Tuo, Hong-Bo~Zhang and Qiping~Su, Phys.\ Rev.\ D {\bf 84}, 123501 (2011); Rong-Gen~Cai, Zhong-Liang~Tuo, Ya-Bo~Wu and Yue-Yue~Zhao, Phys.\ Rev.\ D {\bf 86}, 023511 (2012); M.~Biswas, U.~Debnath, S.~Ghosh and B.~K.~Guha, Eur.\ Phys.\ J.\ C {\bf 79}, 659 (2019); M.~Rezaei, J.~Solà-Peracaula and M.~Malekjani, Mon.\ Not.\ Roy.\ Astron.\ Soc. {\bf 509}, 2593 (2021); H.~Hossienkhani, H.~Yousefi, N.~Azimi and Z.~Zarei, Astrophys.\ Space\ Sci. {\bf 365}, 59 (2020).

\bibitem{higher}
G.~Arciniega, J.~Edelstein and L.~G.~Jaime, Phys.\ Lett.\ B {\bf 802}, 135272 (2020).

\bibitem{saridakiscubic}
C.~Erices, E.~Papantonopoulos and E.~N.~Saridakis, Phys.\ Rev.\ D {\bf 99}, 123527 (2019).

\bibitem{cisternaquartic}
A.~Cisterna, N.~Grandi and J.~Oliva, Phys.\ Lett.\ B {\bf 805}, 135435 (2020).

\bibitem{chemical}
J.~A.~S.~Lima and J.~S.~Alcaniz, Phys.\ Lett.\ B {\bf 600}, 191 (2004); J.~A.~S.~Lima and S.~H.~Pereira, Phys.\ Rev.\ D {\bf 78}, 083504 (2008); S.~H.~Pereira and J.~A.~S.~Lima, Phys.\ Lett.\ B {\bf 669}, 266 (2008).

\bibitem{callen}
H.~B.~Callen, {\it Thermodynamics and an introduction to Thermostatistics}, John Wiley, (1985).

\bibitem{maartens}
R.~Maartens, arXiv:astro-ph/9609119.

\bibitem{reif}
F.~Reif, {\it Fundamentals of Statistical and Thermal Physics}, Waveland Press, (2009).

\bibitem{lepegrandon}
V.~H.~C\'ardenas, D.~Grand\'on and S.~Lepe, Eur.\ Phys.\ J.\ C {\bf 79}, 357 (2019).

\bibitem{interact}
B.~Wang, E.~Abdalla, F.~Atrio-Barandela and D.~Pav\'on, Rept.\ Prog.\ Phys. {\bf 79}, 096901 (2016).

\bibitem{nat1}
E.~M.~Purcell and R.~V.~Pound, Phys.\ Rev. {\bf 81}, 279 (1951).

\bibitem{nat2}
S.~Braun, J.~P.~Ronzheimer, M.~Schreiber, S.~S.~Hodgman, T.~Rom, I.~Bloch and U.~Schneider, Science {\bf 339}, 52 (2013).

\bibitem{saridakis}
E.~N.~Saridakis, P.~F.~Gonz\'alez-D\'\i az and C.~L.~Sig\"uenza, Class.\ Quantum\ Grav. {\bf 26}, 165003 (2009).

\bibitem{relate1}
J.~A.~S.~Lima A.~Del~Popolo and A.~R.~Plastino, Phys.\ Rev.\ D {\bf 100}, 104042 (2019).

\bibitem{relate2}
O.~Klein, Rev.\ Mod.\ Phys. {\bf 21}, 531 (1949).

\bibitem{chemical2}
M.~Cruz, S.~Lepe, S.~D.~Odintsov, Phys.\ Rev.\ D {\bf 98}, 083515 (2018).

\bibitem{nat3}
E.~Abraham and O.~Penrose, Phys.\ Rev.\ E {\bf 95}, 012125 (2017). 

\bibitem{nat4}
J.~P.~P.~Vieira, C.~T.~Byrnes and A.~Lewis, J.\ Cosmol.\ Astropart.\ Phys. \textbf{16}, 060 (2016).

\bibitem{nat5}
P.~F.~Gonz\'alez-D\'\i az and C.~L.~Sig\"uenza, Nucl.\ Phys.\ B {\bf 697}, 363 (2004).

\bibitem{nat6}
D.~Youm, Phys.\ Lett.\ B {\bf 531}, 276 (2002).

\bibitem{apparent}
S.~A.~Hayward, Class. Quantum\ Grav. {\bf 15}, 3147 (1998); P.~Bin\'etruy and A.~Helou, Class. Quantum\ Grav. {\bf 32}, 205006 (2015); D.~Bak and Soo-Jong~Rey, Class.\ Quantum\ Grav. {\bf 17}, L83 (2000). 

\bibitem{cai}
R.~G.~Cai and S.~P.~Kim, J.\ High\ Energy\ Phys. {\bf 2005}, 050 (2005); R.~G.~Cai, Li-Ming~Cao and Ya-Peng~Hu, Class.\ Quantum\ Grav. {\bf 26}, 155018 (2009) 

\bibitem{appa2}
D.~Wenjie-Tian and I.~Booth, Phys.\ Rev.\ D {\bf 92}, 024001 (2015).

\bibitem{sariappa1}
A.~Lymperis and E.~N.~Saridakis, Eur.\ Phys.\ J.\ C {\bf 78}, 993 (2018).

\bibitem{sariappa2}
E.~N.~Saridakis, J.\ Cosmol.\ Astropart.\ Phys. \textbf{07}, 031 (2020).

\bibitem{vagnozzi3}
S.~Vagnozzi, S.~Dhawan, M.~Gerbino, K.~Freese, A.~Goobar and O.~Mena, Phys.\ Rev.\ D {\bf 98}, 083501 (2018).
\end{thebibliography}
\end{document}